# The Intermittent Nature of Leaf Growth Fields

Shahaf Armon[1], Michael Moshe[2] and Eran Sharon[2]

(1) Department of physics of complex systems, Weizmann Institute of Science, Rehovot, Israel

(2) Department of physics, The Hebrew University, Jerusalem, Israel

What are the general principles that allow proper growth of a tissue or an organ? A growing leaf is an example of such a system: it increases its area by orders of magnitude, maintaining a proper (usually flat) shape. How can this be achieved without a central control unit? One would think that a combination of uniform growth with viscoelastic rheology would allow that. Here we show that that the exact opposite process is in action: the natural growth of the leaf surface strongly fluctuates in time and position. By combining high resolution measurements and multi-scale statistical analysis, we suggest a paradigm-change in the way leaf growth is viewed. We measure the in-plane tissue growth of Tobacco leaves in Lagrangian coordinates in 3D and study the statistics of three scalar fields associated with the tensorial growth field: The growth rate, its isotropy and directionality. We identify the governing time and length scales of the fluctuations in the growth rate, and capture abundant switching between local area swelling and shrinking, when measured in high resolution. At lower spatio-temporal resolution the growth rate field becomes smooth. In contrast, the anisotropy field increases over time. Finally, we find significant differences between growth measured during day and night. The growth at night is found to be more intermittent, with shorter correlation lengths and no global directionality. Despite its fluctuative nature, growth fields are not random, thus carry information about growth regulation. Indeed, mechanical analysis shows that a growing leaf can stay flat only if the measured fluctuations are regulated/correlated. Our measurements suggest that the entire statistics of growth fields, and not just their means, should be studied. In particular, the regulation of such fields and the link between their characteristics and the global geometry of a leaf should be studied.

## Introduction

Shaping of soft tissue organs via growth is a highly nontrivial process. It involves biological and chemical processes, and at the same time it generates, and is possibly affected by, physical fields. In living systems, growth occurs within elementary units – the cells. An uncorrelated growth of cells is likely to achieve a geometrical incompatibility, leading to the buildup of internal stresses [1, 2]. The accumulation of such internal stresses may lead to a morphological distortion of the organ [3]. Therefore, cell growth must be regulated in order to generate the precise shape of an organ, weather flat or spatially complex. Currently, it is not well understood how cells growth is controlled to achieve a desired organ shape.

Different genetic and molecular networks underlying growth regulation were proposed and studied. Some genes expressions are known to correlate with dynamic fronts of enhanced growth or growth-arrest, for example *KLUH* [4],[5] and *CIN-TCP* [6, 7]. Other genetic factors are known to enhance/reduce growth rates by delaying or promoting cell maturation, for example *JAGGED* [8] and YABBY and KNOX transcription factors [9, 10]. Hormones and cell-produced materials, like auxin, expansin and pectin products (PMBs), are known to be involved in determining growth rates and growth directionality by loosening cell walls [11–13]. More recently, the mechanical stress field was proposed to be involved in growth regulation at the tissue level. Mechanical effects on cells growth were studied in animals, mainly in Drosophila embryo development [11, 14–19] In plants, it has been shown that a directional stress on plant tissue could affect the orientation of fibers in the cell wall, thus leading to anisotropic growth, correlated with the stress [11, 20]. However, it is not known how these factors, that affect cell mechanics and growth, are coordinated in a way that leads to proper growth of the leaf.

Plant cells, unlike animal cells, do not migrate during development. The cells are surrounded by rigid cell walls that are cemented to the walls of neighboring cells and create a permanent matrix. In plant lateral organs, such as leaves and petals, cells division is stopped at early stages of development and most of the increase in volume occurs via cell expansion (typical 10-100 fold increase in volume and up to a cell diameter of 10-100 microns). The main force that drives cell expansion is the turgor pressure – the osmotic pressure between the exterior and the interior of the cell. This pressure is of order 0.1-1 MPa [11] and can be varied via changes in ions concentration in the cell. The shape and size of the pressurized cells are stabilized by the high stiffness of their cell walls. These visco-elastic walls are made of polymer networks and cell expansion is thought to involve yielding of the cell wall to the internal pressure [21] and deposition of new material from within the cell. Based on this mechanical view of the process, one would expect leaf growth to be smooth and isotropic, as in the swelling of a balloon. In such a scenario, growth regulation is expected to be active in large length scales, balancing long wavelength inhomogeneity in growth.

However, measurements showed that growth is not uniform across a leaf. For example, large variations in cell expansion in different regions of the leaf [22, 23] as well as highly anisotropic cell expansion [8] were measured. Moreover, recent measurements and models indicate that the growth field is heterogenous also in small scales, where neighboring cells can undergo extremely different expansion levels [24–27]. So far there was no systematic study of the growth fluctuations and its statistical properties.

In this work we measure the surface growth of Tobacco (*Nicotiana tabacum*) leaves in high spatial and temporal resolutions. We find that the growth field of these leaves is far from being smooth. In fact, it is a multiscale field, which involves sharp fluctuations at small time and length scales. Only at large scales in time and space the field becomes smooth and well represented by its means. We study the statistics of the relevant quantities, such as the growth rate, isotropy and directionality. The results reveal several new facts about leaf growth. These include the abundance of local shrinkage events in the tissue during growth, the typical time and length scales of growth heterogeneity and the qualitative difference between leaf growth during day and night. These observations imply that in order to understand leaf growth and its regulation, one must study the statistical characteristics of the growth field.

Treating a leaf as a thin plate, its surface growth can be described by a tensorial field - the "growth tensor" - a rank-two tensor with eigen-vectors that locally represent the directions of maximal and minimal growth, and eigen-values that express the growth rates in these principal directions. In order to measure the statistics of such a field, one needs a system that allows repeated measurements of growth in high spatial and temporal resolution and for long durations of time.

**Materials and Methods**

*Data Acquisition*: Our experimental system is designed to measure the local lateral growth tensor on the top surface of leaves growing free of external constraints. The system includes a profilometer (MiniconScan 3000) which provides the surface topography $z(x, y)$ in high resolution (50 $\mu m$ in x-y, 5 $\mu m$ in z) over the entire leaf (typically $3 \times 3\ cm$). The calculations of growth fields were obtained from the combination of periodic profilometer scans and optical images, taken at 15 min intervals (fig.1A-C, SOM 1).

*Data Processing*: Using image processing and PIV (Particle Image Velocimetry) we obtain the displacement field $\vec{d}(u, v)$ between images on a flat square grid $(u, v)$ of resolution $250 \mu m \times 250 \mu m$ (approximately 10 by 10 cells).

We project the 2D grids of sequential measurements, onto the relevant topographical scans, obtaining the 3D grids before and after the growth $u(x, y, z, t), v(x, y, z, t)$. Using these Lagrangian coordinates, that evolve with the leaf surface, we calculate the local growth tensor, $G(u, v)$. For each Lagrangian grid point we calculate the ratio between the surface geometrical metrics in time $t_1$ and $t_2$. The result is a locally defined rank-two tensor (see SOM2) whose eigen vectors represent the directions of maximal and minimal growth. The maximal and minimal local elongations between $t_1$ and $t_2$ are $\sqrt{\lambda_1} - 1, \sqrt{\lambda_2} - 1$ respectively, where $\lambda_1, \lambda_2$ are the eigen values of $G(u, v)$. The local area growth is $AG = \sqrt{\lambda_1 \lambda_2} - 1$ (The normalized change in area of a surface element at $t_1$ and $t_2$). The estimated measurement error in $AG$ is ~0.2% (see SOM3). Growth rates, are obtained by dividing the growth values by $\Delta t = t_2 - t_1$. We define the isotropy to be $I = \frac{\lambda_2}{\lambda_1}$ where $\lambda_1 > \lambda_2$. We define the main growth angle, ɸ, to be the angle between the maximal growth direction and the leaf main vein. For a full description of the local growth, any three independent scalar parameters are sufficient, e.g. ($\lambda_1, \lambda_2, $ɸ).

## Results

A Tobacco (*Nicotiana tabacum*) WT leaf was measured every 15 minutes for 2 days, in which the leaf total area was tripled (28 to 89 mm$^2$). The leaf grew flat and its area increased in ~4% every hour (not shown). The increase in leaf surface area reflects the average of the local growth rate.

Plotting the local surface growth field across the leaf, $AG(u,v)$, during a specific 15 minutes interval (fig. 1D), we find it fluctuates widely. Surprisingly, although the leaf grew during that time interval, i.e. its surface area increased, many spots *shrank* (blue areas in fig.1D) during the 15 minutes interval. Plotting the histogram of the local surface growth during 90 intervals of 15 min each (fig.1E) reveals a surprising picture. The histogram is broad and non-Gaussian. A significant part of it consists of negative values of $AG$ and the distribution's width is much larger than its mean. Looking at the growth along the principal directions (fig.1F) reveals that the growth in these scales is not only non-homogenous but also anisotropic. In addition, the principal directions themselves fluctuate in space.

Once realizing the fluctuative nature of the growth field, we turn to study its statistical properties. We start by increasing the time intervals $\Delta t$ over which growth is calculated. In fig.1G,H we present examples of such growth fields calculated on images separated by 1 and 3 hours. In addition to the trivial increase in the amount of growth with $\Delta t$, the growth fields seem to become smoother. While local shrinkage events (red lines) are common during 15 minutes of growth, they completely disappear at larger time intervals, leading to an apparent smooth growth field. Looking at the total distribution of such $AG$ fields vs $\Delta t$ (fig.2A inset) we see that as $\Delta t$ increases, the centers of the histograms shift to larger positive values and the distribution broadens (see also fig.2B). However, the ratio between the mean and the standard deviation (std) of the distributions *decreases* with $\Delta t$ (fig.2B inset), reflecting the fact that the growth field becomes smoother when measured over longer time intervals. The shape of the histograms is not Gaussian (fig.2A), having stretched exponential tails. Its higher moments, however, are constant in $\Delta t$ (skewness$\cong 1$, kurtosis$\cong 5$). Finally, our measurements indicate that the growth field during night time is more fluctuative than the growth during day time (fig.2B inset).

Next, we study the spatial properties of the growth fields. We coarse grain the growth field in the spatial scale (by decreasing the spatial resolution of the displacement field measurement). The growth grid step, $\Delta x$, is increased from 16 pixels ($240 \mu m$) gradually to 560 pixels (1.44 mm)(fig.2C). As expected, the mean of the distribution is not affected by the coarse graining, but the standard deviation decreases (fig.2D). As a result, the growth fields measured with lower spatial resolution seem much smoother than they actually are. In this sense, measurements of low spatial and temporal resolution give the erroneous impression of a smooth expansion of the leaf surface.

The large fluctuations in the measured growth field are highly relevant to the question of growth regulation. If the fluctuations in growth were not correlated, one would expect a buildup of large internal stresses within the leaf tissue (SOM 4) and possibly its distortion into a 3D configuration via buckling instabilities. The fact that such distortions are not observed indicates that despite its large fluctuations, growth is regulated on a relatively small time and length scales.

The spatial correlation function of the $AG$ provides information about the spatial scales of regulation. Four selected results, obtained for both day and night, are shown in fig.3A. All four correlation functions decay to zero over a finite distance. As expected, the measured decorrelation length increases with coarse graining in time. A distinct result is the big difference between growth during day and night. The decorrelation length at night is significantly shorter than the ones measured for growth during day time (~ 1mm vs. ~5 mm). Anti-correlation was measured for night growth during the shortest measured intervals (15 min.).

Next, we attempt to identify a *typical time scale* that governs the fluctuations in the area growth fields. We measure growth in $\Delta t$ =15min time intervals over long duration of time (8 hours). The measurements are converted to Lagrangian coordinates and $AG$ is evaluated in grid points that are continuously "drifting" with growth (i.e. locations in the growing leaf are fixed for the entire 8 hours). For each such Lagrangian grid point, we perform Fourier transform of the signal $AG(t)$ with $\Delta t$=15min. We then average the absolute value of the result over all grid points, obtaining $AG(\omega)$. A clear peak in $AG(\omega)$ is seen at $\omega = 0.14\ [min^{-1}]$ (fig.3B), which indicates a typical time scale of ~45minutes. Therefore, the growth rate in a typical fixed location on a leaf fluctuates with a characteristic time scale.

We now switch our interest to the directional properties of the tensorial growth field. The distributions of maximal and minimal growth values for $\Delta t = 15 min$ are presented in fig.4A. The histograms of the growth anisotropy, $I = \frac{\lambda_2}{\lambda_1}$, show an increase in the mean value of anisotropy as functions of $\Delta t$ (fig.4B) and does not decay with $\Delta t$, as expected if the orientation would be random (fig.4C). Finally, we plot a histogram of the growth angle, $\phi$ (the angle between maximal growth and the leaf main vein) (fig.4D). The histogram shows that during the day the preferred orientation of the local maximal growth is perpendicular to the main vein, while during night there is no preferred directionality of growth.

**Discussion**:

Our measurements suggest a new view on leaf growth. The growth is not smooth in time and space as could be concluded from low resolution measurements, or from data that is averaged over many leaves. Instead, growth fields are highly fluctuative, and therefore should be analyzed using tools commonly applied in stochastic dynamics of fluids, solids and active matter. We found abundance of local tissue shrinkage during growth. Though the surface area of large sections of a leaf increases monotonically (in a typical rate of ~4%/hour), the growth in small regions, of ~100 cells, oscillates between swelling and shrinking. These oscillations turn out to be part of the natural growth process. They are characterized by a typical time scale (~45 min) and are correlated over short distances of order millimeters. The higher is the measurement resolution in time and scale, we show larger fluctuations. It is plausible that increasing resolution in time and space (to a single cell level) would reveal even rougher fields. Decreasing resolution, to 1 hour or a few mm, the ratio of std to mean saturates to a low value (~0.5, fig 2B,D). Note, that the speed scale of reaching this saturation (+-2% area in 1 hour) is similar to the average speed of growth itself (5%area/hour), meaning, that growth is inseparable from the process of relaxation/homogenization. The anisotropy of growth fluctuates as well, however contrarily to the area growth, it is not averaged out in time.

One immediately notes that these new observations are highly relevant to the question of growth regulation. How can such a highly variable growth field lead to smooth growth of a flat leaf? What causes all these large fluctuations to be properly averaged? The non-gaussian growth distribution (fig. 1E,fig2A,C), and the persistent anisotropy (fig.4C) indicate that indeed, though rough in time and space – the growth field is far from being random.

If growth was a random process, i.e. growth parameters $(\lambda_1, \lambda_2, \phi)$ were chosen randomly from some distribution at each grid point independently, the std of the spatial distribution of the total growth would grow like $\sqrt{t}$. Using our measured distributions and a random growth model (SOM 4), we estimate a typical stress developing within 15 minutes on a flat unit area of 250x250 micron to be in the order of ~0.5Mpa. Using the distributions of growth in two hours, and assuming no spatial correlations, the stress estimation on the same unit area would exceeds 1MPa. Such stress would have led to significant buckling of the surface and distortion of the leaf. In reality, the leaf grows flat, which indicates that some mechanism controls growth over small time and length scales.

Finally, we show that the growth characteristics vary between day and night. The growth during day time is smoother and more coherent. The correlation lengths in the growth fields are larger during the day, and the direction of growth is oriented roughly perpendicularly to the main vein. Growth at night is characterized by a lack of global orientation, higher fluctuations and short correlation lengths. It is possible that these two different modes of growth are combined to generate the globally coherent growth: most of the active growth takes place during the day, and the accumulated internal stresses are released during the night.

The results presented in this work call for further experimental, as well as theoretical and numerical, studies. Though we observed qualitatively similar results in measurements of many Tobbacco and Arabidopsis leaves, measurements in other species are needed in order to determine how general our observations are. In addition, the stochastic approach should be implemented on measurements at the cellular scale. Finally, studies of mathematical and physical nature are needed in order to better characterize the relation between statistics of growth, mechanics of the leaf, and its final global shape.

**Acknowledgements**:

We thank Prof Naomi Ori for providing the plants and the advice for their growth; Oded Ben David for his help with the experimental setup, and Hillel Aharoni for useful theoretical discussions.

**References**

1. Efrati E, Sharon E, Kupferman R (2009) Elastic theory of unconstrained non-Euclidean plates. J Mech Phys Solids 57:762–775

2. Wang C-C (1968) On the Geometric Structure of Simple Bodies, a Mathematical Foundation for the Theory of Continuous Distributions of Dislocations. In: Mechanics of Generalized Continua. Springer Berlin Heidelberg, Berlin, Heidelberg, pp 247–250

3. Goriely A, Ben Amar M (2005) Differential Growth and Instability in Elastic Shells. Phys Rev Lett 94:198103. https://doi.org/10.1103/PhysRevLett.94.198103

4. Anastasiou E, Kenz S, Gerstung M, et al (2007) Control of Plant Organ Size by KLUH/CYP78A5-Dependent Intercellular Signaling. Dev Cell 13:843–856

5. Kazama T, Ichihashi Y, Murata S, Tsukaya H (2010) The mechanism of cell cycle arrest front progression explained by a KLUH/CYP78A5-dependent mobile growth factor in developing leaves of Arabidopsis thaliana. Plant Cell Physiol 51:1046–54. https://doi.org/10.1093/pcp/pcq051

6. Nath U, Crawford BCW, Carpenter R, Coen E (2003) Genetic control of surface curvature. Science 299:1404–7. https://doi.org/10.1126/science.1079354

7. Palatnik JF, Allen E, Wu X, et al (2003) Control of leaf morphogenesis by microRNAs. Nature 425:257–63. https://doi.org/10.1038/nature01958

8. Sauret-Güeto S, Schiessl K, Bangham A, et al (2013) JAGGED controls Arabidopsis petal growth and shape by interacting with a divergent polarity field. PLoS Biol 11:e1001550. https://doi.org/10.1371/journal.pbio.1001550

9. Burko Y, Ori N (2013) The tomato leaf as a model system for organogenesis. Methods Mol Biol 959:1–19. https://doi.org/10.1007/978-1-62703-221-6_1

10. Ha CM, Jun JH, Fletcher JC (2010) Control of Arabidopsis leaf morphogenesis through regulation of the YABBY and KNOX families of transcription factors. Genetics 186:197–206. https://doi.org/10.1534/genetics.110.118703

11. Mirabet V, Das P, Boudaoud A, Hamant O (2011) The role of mechanical forces in plant morphogenesis. Annu Rev Plant Biol 62:365–85. https://doi.org/10.1146/annurev-arplant-042110-103852

12. Hasson A, Blein T, Laufs P (2010) Leaving the meristem behind: The genetic and molecular control of leaf patterning and morphogenesis. C R Biol 333:350–360

13. Efroni I, Eshed Y, Lifschitz E (2010) Morphogenesis of simple and compound leaves: a critical review. Plant Cell 22:1019–32. https://doi.org/10.1105/tpc.109.073601

14. Labouesse M, Farge E (2011) Mechanotransduction in Development. In: Current Topics in Developmental Biology. pp 243–265

15. Chiou KK, Hufnagel L, Shraiman BI (2012) Mechanical stress inference for two dimensional cell arrays. PLoS Comput Biol 8:e1002512. https://doi.org/10.1371/journal.pcbi.1002512

16. Schluck T, Nienhaus U, Aegerter-Wilmsen T, Aegerter CM (2013) Mechanical Control of Organ Size in the Development of the Drosophila Wing Disc. PLoS One 8:e76171. https://doi.org/10.1371/journal.pone.0076171

17. Irvine KD, Shraiman BI (2017) Mechanical control of growth: ideas, facts and challenges. Development 144:4238–4248. https://doi.org/10.1242/dev.151902

18. Chiou KK, Hufnagel L, Shraiman BI Mechanical stress inference for two dimensional cell arrays. PLoS Comput Biol 8:e1002512

19. LeGoff L, Lecuit T (2016) Mechanical Forces and Growth in Animal Tissues. Cold Spring Harb Perspect Biol 8:a019232. https://doi.org/10.1101/cshperspect.a019232

20. Hamant O, Heisler MG, Jönsson H, et al (2008) Developmental Patterning by Mechanical Signals in Arabidopsis. Science (80- ) 322:1650–1655. https://doi.org/10.1126/SCIENCE.1165594

21. Lockhart JA (1965) An analysis of irreversible plant cell elongation. J Theor Biol 8:264–275

22. Wiese A, Christ MM, Virnich O, et al (2007) Spatio-temporal leaf growth patterns of Arabidopsis thaliana and evidence for sugar control of the diel leaf growth cycle. New Phytol 174:752–61. https://doi.org/10.1111/j.1469-8137.2007.02053.x

23. Remmler L, Rolland-Lagan A-G (2012) Computational method for quantifying growth patterns at the adaxial leaf surface in three dimensions. Plant Physiol 159:27–39. https://doi.org/10.1104/pp.112.194662

24. Elsner J, Michalski M, Kwiatkowska D (2012) Spatiotemporal variation of leaf epidermal cell growth: a quantitative analysis of Arabidopsis thaliana wild-type and triple cyclinD3 mutant plants. Ann Bot 109:897–910. https://doi.org/10.1093/aob/mcs005

25. Sampathkumar A, Krupinski P, Wightman R, et al (2014) Subcellular and supracellular mechanical stress prescribes cytoskeleton behavior in Arabidopsis cotyledon pavement cells. 3:1967. https://doi.org/10.7554/eLife.01967

26. Fruleux A, Boudaoud A (2019) Modulation of tissue growth heterogeneity by responses to mechanical stress. Proc Natl Acad Sci U S A 116:1940–1945. https://doi.org/10.1073/pnas.1815342116

27. Hong L, Dumond M, Tsugawa S, et al (2016) Variable Cell Growth Yields Reproducible OrganDevelopment through Spatiotemporal Averaging. Dev Cell 38:15–32. https://doi.org/10.1016/j.devcel.2016.06.016

28. Efrati E, Sharon E, Kupferman R (2009) Buckling transition and boundary layer in non-Euclidean plates. Phys Rev E 80:016602. https://doi.org/10.1103/PhysRevE.80.016602

29. Sahaf M, Sharon E (2016) The rheology of a growing leaf: stress-induced changes in the mechanical properties of leaves. J Exp Bot 67:5509–5515. https://doi.org/10.1093/jxb/erw316

**Supplementary material for**

**"The Intermittent Nature of Leaf Growth Fields"**

1. **Leaf surface measurements**

In order to measure the growth of a leaf lamina top surface, a living plant is placed in its pot on a moving stage.

The 3D leaf surface is measured using an optical profilometer (MiniconScan 3000, by Optimet), which provides the surface height $z(x, y)$ using conical holography. We measure the leaf in high resolution scans (50 $\mu$m in x-y, 5 $\mu$m in z) on a large field of view (up to $3 \times 3$ cm). The measured surface is smoothed using a smoothing spline algorithm (Matlab7 Toolbox) in order to eliminate trichomes and other small scale height fluctuations. All leaf scans were programmed to last less than 10 minutes, in order to neglect shape changes during the scans.

Adjacent to the scans, a camera is taking sequential photos of the leaf from top view. The camera (Luminera Lw575) is set at a distance of 200 cm from the plant in order to reduce growth artifacts, due to the leaf movement towards the lens. This way, we lower the effect to less than 0.5% in area growth (see fig.S1). The remaining error is fixed in post calculation by a calibration.

The chosen leaf is periodically scanned and then imaged every 15 minutes (Microsoft Task Scheduler, Stepnet controller by Copley, Visual Basic) for several days. Lighting is controlled computationally to induce short-day conditions (8 hours of light per day, 6500k). Throughout the paper, lighted hours are called 'day time' and no-light hours are called 'night'. During the night, a flash of light is scheduled for the picture taking. The scanned leaf is kept free of any external constrains. The soil humidity is kept high during all times.

The system can measure the shape and growth of complex leaves in 3D. In this work, however, we measure Tobacco WT plants, which have rather flat leaves, which are oriented relatively horizontally.

The measured leaf presented in this work was leaf #6 in the order of growth.

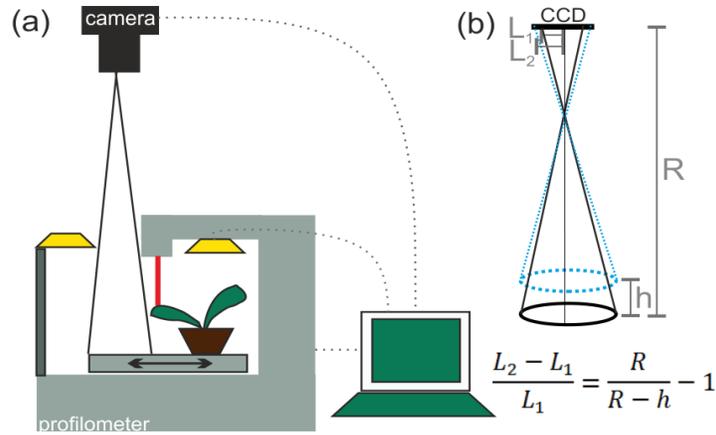

**Fig S1:** The leaf growth measuring system: (a) A pot with a living plant is put on the x-y stage of the Profilometer. The stage is moving automatically according to a preset plan, so the leaf is scanned and imaged in constant time interval. A camera is set on the ceiling to minimize growth artifacts. The lighting conditions (lamps marked in yellow) are also controlled computationally. When a measurement is taken during the night, light flashes to enable the image taking. (b) A rough sketch of the artifact surface "growth" due to the leaf rising up towards the camera. By lowering the ratio $\frac{h}{R}$ we reduce this artifact.

## 2. Calculation of the growth tensor

In order to measure a growth between time $t_1$ and $t_2$, we perform high-pass filter (Matlab7) on the two relevant images, in order to enhance the leaf features like veins, trichomes and other fine details (see example image in fig.1C in main text). On the resulting two images we perform PIV (Particle Image Velocimetry, Matlab7 2012a): The algorithm calculates the displacement field $\vec{d}(u,v)$ between these images on a pre defined square grid. For each grid square in the first image, the algorithm finds a square in the second image with maximal correlation to it. The vectors connecting between the centers of squares in the first and second images create the 2D displacement field. Our measurements were taken on a $16 \times 16$ pixels grid, which is about $250\mu m \times 250\mu m$ (about 10 by 10 cells).

The next step includes registration between the leaf image and the leaf 3D scan. This is done once per experiment, using manual pointing on 3 obvious leaf features that appear both in the profilometer scan and in the optical image. By accurately projecting the PIV result onto the topographic scans, we obtain a 3D grid $u(x, y, z, t)$, $v(x, y, z, t)$ before and after the growth. Using these Lagrangian coordinates that are on the leaf $(u, v)$, we calculate the surface metric at each grid point at time t1 and t2:

$$a_1 = \begin{pmatrix} \frac{\partial^2 z_1}{\partial u^2} & \frac{\partial^2 z_1}{\partial u \partial v} \\ \frac{\partial^2 z_1}{\partial v \partial u} & \frac{\partial^2 z_1}{\partial v^2} \end{pmatrix}; a_2 = \begin{pmatrix} \frac{\partial^2 z_2}{\partial u^2} & \frac{\partial^2 z_2}{\partial u \partial v} \\ \frac{\partial^2 z_2}{\partial v \partial u} & \frac{\partial^2 z_2}{\partial v^2} \end{pmatrix}$$

We define the growth tensor to be the ratio between these two metrics. This ratio is calculated locally in the following way: We find the local coordinate transformation $J_{ij}$ after which the initial metric $a_1$ is the unit matrix and the final metric $a_2$ is diagonal. The procedure can always be done, since a metric is a real, symmetric matrix, hence can always be diagonalized by an orthogonal transformation (rotation). The growth tensor, $G(u,v)$, is the ratio between the metrics $a_1$ and $a_2$ written after the coordinate transformation J. The algorithm is formally described as follows:

$$A^T a_1 A = \begin{pmatrix} 1 & 0 \\ 0 & 1 \end{pmatrix}$$

$$O^T A^T a_2 A O = J^T a_2 J = \begin{pmatrix} \lambda_1 & 0 \\ 0 & \lambda_2 \end{pmatrix}$$

$$G(u,v) \equiv \frac{J^T a_2 J}{J^T a_1 J} = \begin{pmatrix} \lambda_1 & 0 \\ 0 & \lambda_2 \end{pmatrix}$$

where A is a combination of a rotation and a non-isotropic swell, O is a pure rotation and J is an integration of them both $J = AO$. The growth tensor, $G(u,v)$, is in fact the metric after growth, $a_2$, written in coordinates in which the initial metric, $a_1$, is unity. The growth tensor eigen vectors (column vectors of J, written in $(u,v)$ coordinates) are the directions of maximal and minimal growth. The maximal and minimal elongations between $t_1$ and $t_2$ are $\sqrt{\lambda_1} - 1, \sqrt{\lambda_2} - 1$ respectively. The elongations in any other direction can be revealed by further tensorial coordinate transformations. The local area growth is $= \sqrt{\lambda_1 \lambda_2} - 1$, which is the ratio between the local surface area at $t_1$ and $t_2$. We define the Isotropy to be $I = \frac{\lambda_2}{\lambda_1}$ where $\lambda_1 > \lambda_2$. The main growth angle, ϕ, is defined as the angle between maximal growth direction and the leaf main vein. A full description of the local growth is given by any three independent scalar parameters, e.g. $(\lambda_1, \lambda_2, \phi)$.

Eventually, we locally "correct" the growth tensor by performing a calibration (once per camera setting) to measure the artificial "growth" effect due to the leaf rising up towards the camera. All post-processing methods were implemented in Matlab7.

### 3. Error estimations

The presented leaf growth measurements were preceded by analysis of dozens of leaves, both of Tobacco and Arabidopsis. In all cases, the fluctuative nature of growth was observed.

Our main measurement errors result from the PIV process, due to a changing light distribution on the leaf, or relative movement of trichomes. Another source for error is the registration between the leaf scan and its image. We tested our measuring system on non-growing objects as well as on inflated balloons and leaves. Deliberate distortions in light conditions and table stability created an error of no more than 0.002 in AG. Deliberate registration errors yielded 0.003 error in AG. Both are negligible compared to our results.

4. **Random growth models, estimating strains and stresses**

As a basal random growth model, we assume the local area growth is chosen randomly from some distribution at each grid point independently. In such a scenario, coarse graining in time means a summation of such random variables, hence increase in the spatial mean proportionally to $\Delta t$ and increase in the spatial std proportionally to $\sqrt{\Delta t}$. coarse graining in space means averaging on groups of such random variables, hence the spatial mean stays constant while the spatial std goes like $1/\sqrt{\Delta x}$.

In a more elaborate random growth model, three growth parameters $(\lambda_1, \lambda_2, \phi)$ are chosen randomly from some distribution at each grid point independently. In this case calculation is not trivial. Due to the random growth directions, the tensorial components of $AG$ mix at each time step. However, it is still clear that both the spatial mean and std of the area growth should increase with $\Delta t$. The spatial mean of the anisotropy should decreases with $\Delta t$ while its std is increasing (fig.4 main text) ,i.e. the anisotropy and its variation increases.

For an estimation of the stress and strain that would develop in such a growing leaf, we use the measured properties of the principal growth distributions in 15 minutes depicted in fig.4A and assume an even distribution of $\phi$, as roughly seen in fig.4D. Using this data we estimate a lower bound for the stress, since sharper fluctuations of growth may exist in smaller scales below our measurement resolution. We denote

$$\langle m \rangle = \int m \, f(m) dm = 0$$

$$\langle M \rangle = \int M \, g(M) dM = 0$$

$$\sqrt{\langle m^2 \rangle} = \sqrt{\int m^2 \, f(m) dm} = l_m$$

$$\sqrt{\langle M^2 \rangle} = \sqrt{\int M^2 \, g(M) dM} = l_M$$

Where f(m) and g(m) are the distribution functions of the minimal and maximal growths respectively, <m> and <M> are their mean, $l_m$ and $l_M$ are their rms values. Examples of f and g for $\Delta t = 15$min is shown in fig.4A in the main text.

We calculate the dilation stress, p, and shear stress, $\sigma$:

$$\sqrt{\langle p^2 \rangle} = \frac{(2\nu + 1)Y}{1 - \nu^2} \sqrt{l_m^2 + l_M^2}$$

$$\sqrt{\langle \sigma^2 \rangle} \sim = \frac{Y}{2\sqrt{2}(1-v^2)} \sqrt{l_m^2 + l_M^2}$$

where Y is Young's modulus and ν is Poisson's ratio of the leaf [28].

Estimating the mechanical coefficients - Young's modulus 20MPa (estimated from [29]) and Poisson's ratio 0.5, we obtain that a typical stress developing within 15 minutes on a flat unit area of 250x250 micron will be in the order of ~0.5Mpa. Using the same random growth model, this time with the measured distributions of growth in two hours (not shown), we obtain a stress estimation on the same unit area that exceed 1MPa. The meaning is that the measured growth distribution but with no spatial correlations would yield stresses in the order of turgor pressure. Note, that our measurements provide only the actual growth of the surface, which by definition, is compatible. It is likely that the fluctuations in the "attempted growth", which generate the internal stresses, are even larger.

References -SI

1. Efrati E, Sharon E, Kupferman R (2009) Elastic theory of unconstrained non-Euclidean plates. J Mech Phys Solids 57:762–775

2. Wang C-C (1968) On the Geometric Structure of Simple Bodies, a Mathematical Foundation for the Theory of Continuous Distributions of Dislocations. In: Mechanics of Generalized Continua. Springer Berlin Heidelberg, Berlin, Heidelberg, pp 247–250

3. Goriely A, Ben Amar M (2005) Differential Growth and Instability in Elastic Shells. Phys Rev Lett 94:198103. https://doi.org/10.1103/PhysRevLett.94.198103

4. Anastasiou E, Kenz S, Gerstung M, et al (2007) Control of Plant Organ Size by KLUH/CYP78A5-Dependent Intercellular Signaling. Dev Cell 13:843–856

5. Kazama T, Ichihashi Y, Murata S, Tsukaya H (2010) The mechanism of cell cycle arrest front progression explained by a KLUH/CYP78A5-dependent mobile growth factor in developing leaves of Arabidopsis thaliana. Plant Cell Physiol 51:1046–54. https://doi.org/10.1093/pcp/pcq051

6. Nath U, Crawford BCW, Carpenter R, Coen E (2003) Genetic control of surface curvature. Science 299:1404–7. https://doi.org/10.1126/science.1079354

7. Palatnik JF, Allen E, Wu X, et al (2003) Control of leaf morphogenesis by microRNAs. Nature 425:257–63. https://doi.org/10.1038/nature01958

8. Sauret-Güeto S, Schiessl K, Bangham A, et al (2013) JAGGED controls Arabidopsis petal growth and shape by interacting with a divergent polarity field. PLoS Biol 11:e1001550. https://doi.org/10.1371/journal.pbio.1001550

9. Burko Y, Ori N (2013) The tomato leaf as a model system for organogenesis. Methods Mol Biol 959:1–19. https://doi.org/10.1007/978-1-62703-221-6_1

10. Ha CM, Jun JH, Fletcher JC (2010) Control of Arabidopsis leaf morphogenesis through regulation of the YABBY and KNOX families of transcription factors. Genetics 186:197–206. https://doi.org/10.1534/genetics.110.118703

11. Mirabet V, Das P, Boudaoud A, Hamant O (2011) The role of mechanical forces in plant morphogenesis. Annu Rev Plant Biol 62:365–85. https://doi.org/10.1146/annurev-arplant-042110-103852

12. Hasson A, Blein T, Laufs P (2010) Leaving the meristem behind: The genetic and molecular control of leaf patterning and morphogenesis. C R Biol 333:350–360

13. Efroni I, Eshed Y, Lifschitz E (2010) Morphogenesis of simple and compound leaves: a critical review. Plant Cell 22:1019–32. https://doi.org/10.1105/tpc.109.073601

14. Labouesse M, Farge E (2011) Mechanotransduction in Development. In: Current Topics in

Developmental Biology. pp 243–265

15. Chiou KK, Hufnagel L, Shraiman BI (2012) Mechanical stress inference for two dimensional cell arrays. PLoS Comput Biol 8:e1002512. https://doi.org/10.1371/journal.pcbi.1002512

16. Schluck T, Nienhaus U, Aegerter-Wilmsen T, Aegerter CM (2013) Mechanical Control of Organ Size in the Development of the Drosophila Wing Disc. PLoS One 8:e76171. https://doi.org/10.1371/journal.pone.0076171

17. Irvine KD, Shraiman BI (2017) Mechanical control of growth: ideas, facts and challenges. Development 144:4238–4248. https://doi.org/10.1242/dev.151902

18. Chiou KK, Hufnagel L, Shraiman BI Mechanical stress inference for two dimensional cell arrays. PLoS Comput Biol 8:e1002512

19. LeGoff L, Lecuit T (2016) Mechanical Forces and Growth in Animal Tissues. Cold Spring Harb Perspect Biol 8:a019232. https://doi.org/10.1101/cshperspect.a019232

20. Hamant O, Heisler MG, Jönsson H, et al (2008) Developmental Patterning by Mechanical Signals in Arabidopsis. Science (80- ) 322:1650–1655. https://doi.org/10.1126/SCIENCE.1165594

21. Lockhart JA (1965) An analysis of irreversible plant cell elongation. J Theor Biol 8:264–275

22. Wiese A, Christ MM, Virnich O, et al (2007) Spatio-temporal leaf growth patterns of Arabidopsis thaliana and evidence for sugar control of the diel leaf growth cycle. New Phytol 174:752–61. https://doi.org/10.1111/j.1469-8137.2007.02053.x

23. Remmler L, Rolland-Lagan A-G (2012) Computational method for quantifying growth patterns at the adaxial leaf surface in three dimensions. Plant Physiol 159:27–39. https://doi.org/10.1104/pp.112.194662

24. Elsner J, Michalski M, Kwiatkowska D (2012) Spatiotemporal variation of leaf epidermal cell growth: a quantitative analysis of Arabidopsis thaliana wild-type and triple cyclinD3 mutant plants. Ann Bot 109:897–910. https://doi.org/10.1093/aob/mcs005

25. Sampathkumar A, Krupinski P, Wightman R, et al (2014) Subcellular and supracellular mechanical stress prescribes cytoskeleton behavior in Arabidopsis cotyledon pavement cells. 3:1967. https://doi.org/10.7554/eLife.01967

26. Fruleux A, Boudaoud A (2019) Modulation of tissue growth heterogeneity by responses to mechanical stress. Proc Natl Acad Sci U S A 116:1940–1945. https://doi.org/10.1073/pnas.1815342116

27. Hong L, Dumond M, Tsugawa S, et al (2016) Variable Cell Growth Yields Reproducible OrganDevelopment through Spatiotemporal Averaging. Dev Cell 38:15–32. https://doi.org/10.1016/j.devcel.2016.06.016

28. Efrati E, Sharon E, Kupferman R (2009) Buckling transition and boundary layer in non-Euclidean plates. Phys Rev E 80:016602. https://doi.org/10.1103/PhysRevE.80.016602

29. Sahaf M, Sharon E (2016) The rheology of a growing leaf: stress-induced changes in the mechanical properties of leaves. J Exp Bot 67:5509–5515. https://doi.org/10.1093/jxb/erw316

**Fig 1:**

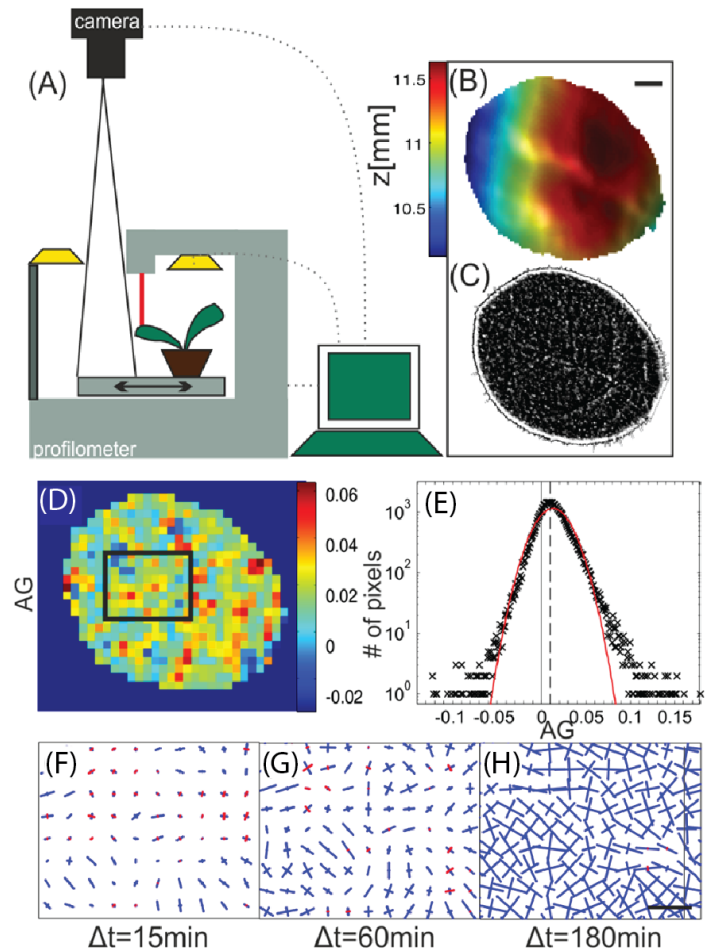

**Fig 1:** The leaf growth measuring system and example dynamics of local growth: (A) A pot with a living plant is placed on the x-y stage of a profilometer. The leaf is automatically scanned and then optically imaged, in constant time intervals. The camera is set at 200cm distance from the leaf plane, in order to minimize optical artifacts due to vertical growth. (B) A typical 3D topography of a Tobbaco WT leaf (after smoothing). (C) An optical image of the leaf in B, after high-pass filtration. The small-scale features are used as tracers for the PIV algorithm. Scale bar: 1mm. (D) An example of a specific area growth field, $AG$, within $\Delta t = 15min$. Resolution: $250 \times 250$ micron (E) A histogram of all area growth events of $\Delta t = 15min$ during 2 days of measurement. Most probable value is ~1% (dashed line). A Gaussian distribution with the same mean and std is depicted in solid red line. (F-H) Principal growth measurements, taken from the same leaf area (marked with a rectangle in D) with different time intervals (indicated). At each point, the maximal and minimal growth vectors are depicted. The vectors are oriented along the principal directions and their lengths indicate growth values $\sqrt{\lambda_1} - 1$, $\sqrt{\lambda_2} - 1$ (scaled by the same arbitrary factor in all panels). Blue lines represent positive growth, red lines represent shrinkage. Scale bar: 0.5 mm.

**Fig 2:**

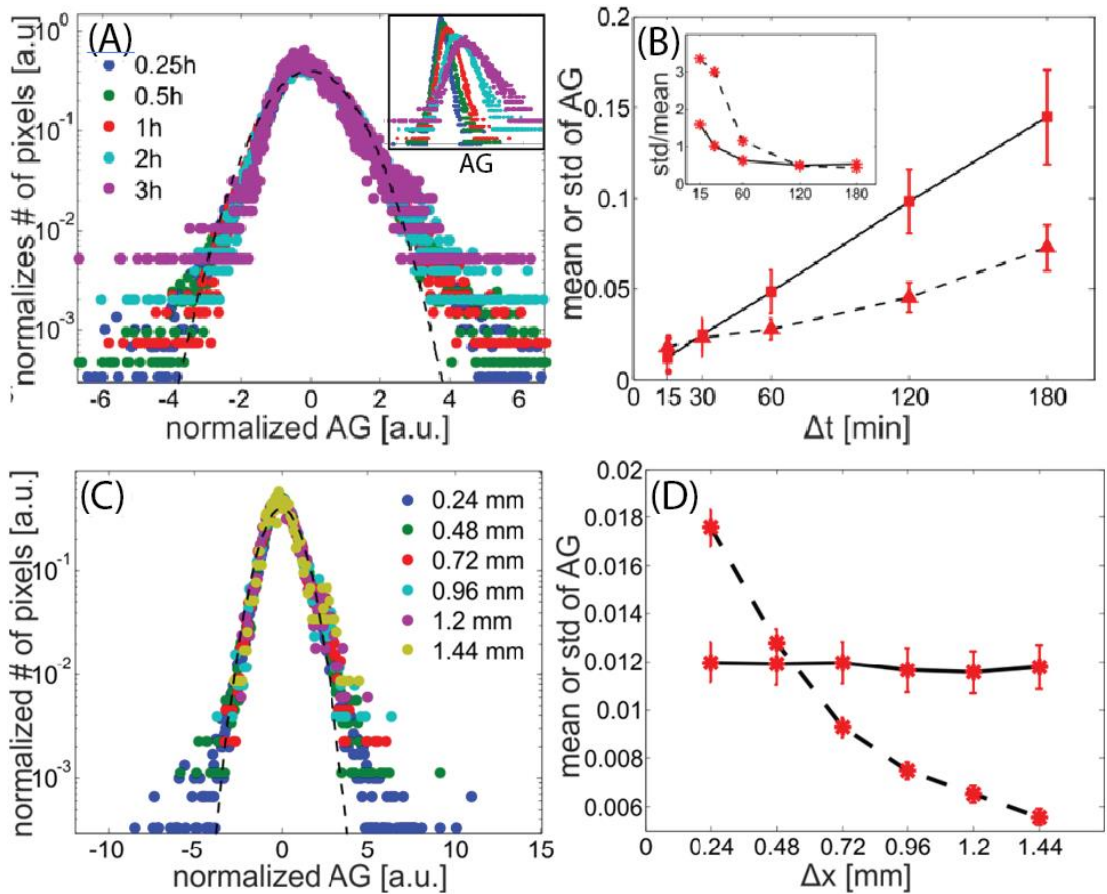

**Fig 2**: Coarse graining in time and space of area-growth fields. (A) Histograms of $AG$ for different $\Delta t$ (indicated in the figure), normalized to have the same std and zero mean. Inset: the same histograms before the normalization. (B) The mean (solid line) and std (dashed line) as functions of $\Delta t$. Inset: the ratio $\frac{std}{mean}$ for measurements that were taken during day time (solid line) and during night time (dashed line). (C) Histograms of $AG$ fields of $\Delta t = 15 min$ calculated for different $\Delta x$ (indicated in the figure). The histograms are normalized to have the same area and zero mean. (D) The mean (solid line) and std (dashed line) of the histograms in C, as functions of $\Delta x$.

**Fig3:**

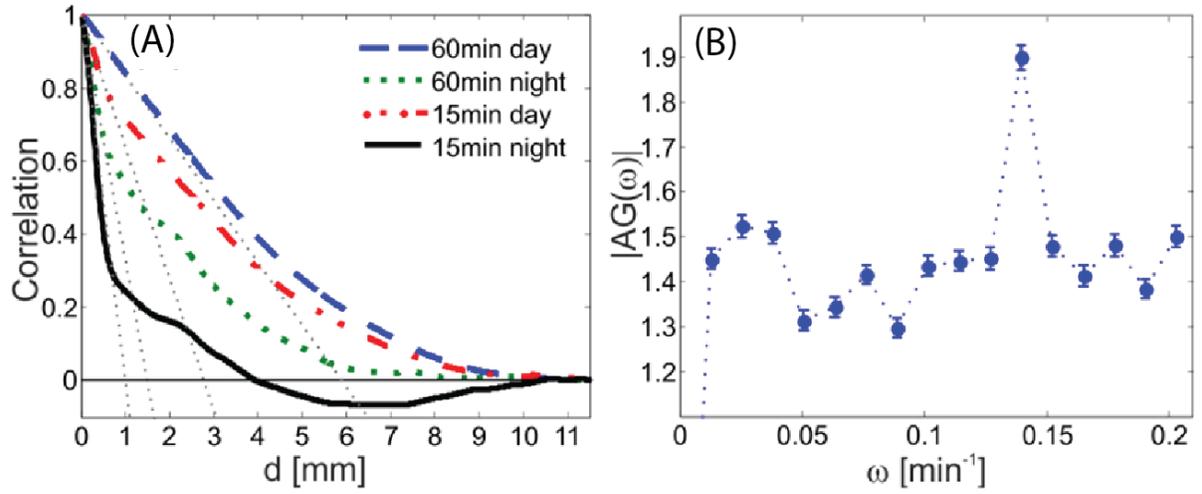

**Fig 3**: Characteristic scales of the area growth field. (A) Correlation between $AG$ values as a function of the distance between measured values, $d$. Data is calculated for four different populations. (B) Mean Fourier transform in time over all signals $AG(t)$ taken in Lagrangian coordinates.

**Fig 4:**

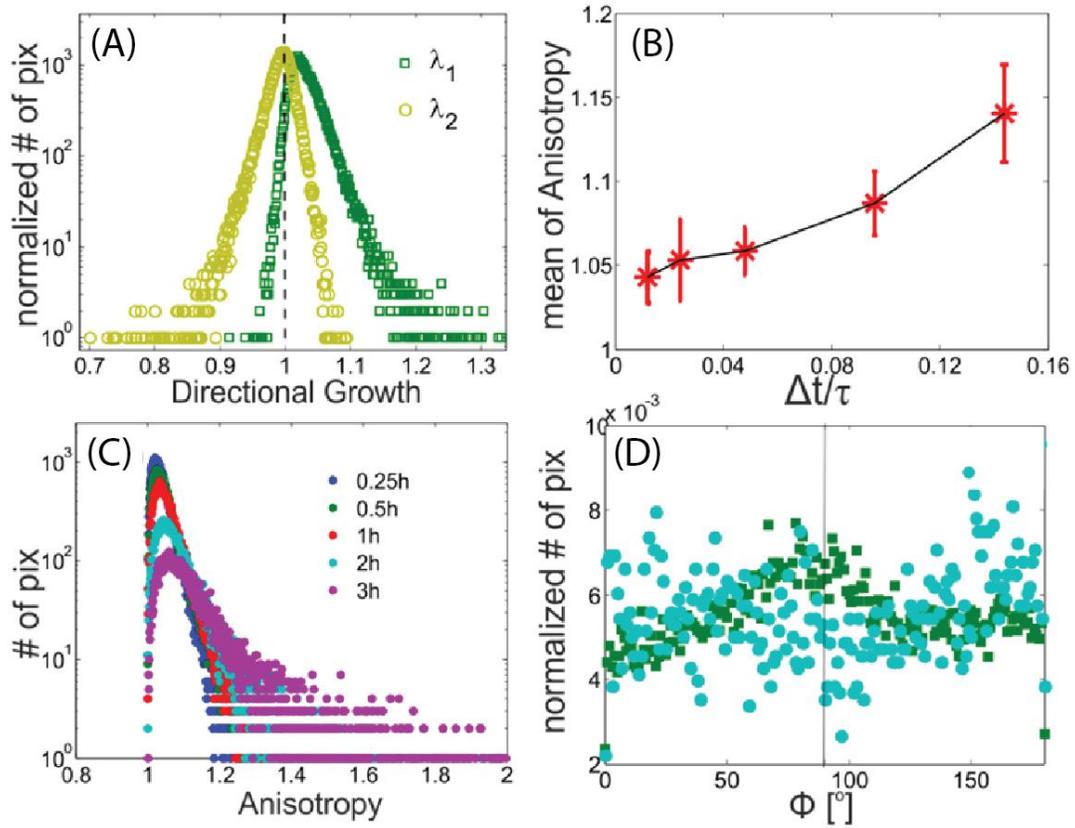

**Fig 4:** Directionality of the growth fields. (A) A histogram of the maximal and minimal directional growths $(\lambda_1, \lambda_2)$ for $\Delta t$=15 min. (B) Histograms of the anisotropy distributions for different $\Delta t$. (C) The means of the histograms in b. (D) A histogram of growth directionality (angle between maximal growth direction and the leaf main vein) for $\Delta t = 15 min$, during the day (green squares) and the night (cyan circles).